\documentclass[10pt,letterpaper]{article}
\usepackage[top=0.85in,left=2.75in,footskip=0.75in]{geometry}

\usepackage{amsmath,amssymb}

\usepackage{changepage}

\usepackage{textcomp,marvosym}

\usepackage{cite}

\usepackage[section]{placeins} 

\usepackage{nameref,hyperref}

\usepackage[right]{lineno}
\newif\ifplossubmission
\plossubmissionfalse

\usepackage[nopatch=eqnum]{microtype}
\DisableLigatures[f]{encoding = *, family = * }

\usepackage[table]{xcolor}

\usepackage{array}

\newcolumntype{+}{!{\vrule width 2pt}}

\newlength\savedwidth

\raggedright
\usepackage[aboveskip=1pt,labelfont=bf,labelsep=period,justification=raggedright,singlelinecheck=off]{caption}

\makeatletter
\renewcommand{\@biblabel}[1]{\quad#1.}
\makeatother

\usepackage{graphicx}
\graphicspath{{./figures}}

\usepackage{lastpage,fancyhdr,graphicx}
\usepackage{epstopdf}
\fancyheadoffset[L]{2.25in}
\fancyfootoffset[L]{2.25in}
\begin{document}
\vspace*{0.2in}

\begin{flushleft}
{\Large
\textbf\newline{PesTwin: A modular agent-based framework for pest and vector population control} % Please use "sentence case" for title and headings (capitalize only the first word in a title (or heading), the first word in a subtitle (or subheading), and any proper nouns).
}
\newline
\\
Andrea De Antoni\textsuperscript{1,2 *},
Giovanni Iacca\textsuperscript{1},
Andrea Pugliese\textsuperscript{1},
Anna Strampelli\textsuperscript{2},
Gerard Terradas\textsuperscript{2},
Matteo Rucco\textsuperscript{2},
Andrew Hammond\textsuperscript{2, *}
\\
\bigskip
\textbf{1} University of Trento, Trento, Italy
\\
\textbf{2} Biocentis S.r.l., Milano, Italy
\\
\bigskip

% Use the asterisk to denote corresponding authorship and provide email address in note below.
* andrea.deantoni-1@unitn.it, andrew.hammond@biocentis.com 

\end{flushleft}
\section*{Abstract}

Species-specific pest and vector control strategies, including the sterile insect technique, Wolbachia-based interventions, and genetic control technologies, offer powerful alternatives to broad-spectrum chemical control, with applications ranging from targeted crop protection to large-scale disease control. Among these, genetic control technologies are advancing particularly rapidly, but the pace of technological development is increasingly outstripping the modelling tools needed to predict outcomes, guide technology design and its implementation, compare alternative strategies across different use settings, and support regulatory and operational decision-making.

Here we present PesTwin, a flexible agent-based modelling framework for simulating genetic control technologies across species, ecological settings, and deployment strategies within a common computational environment. By representing individual insects as autonomous agents carrying explicit genetic information and interacting within spatially structured environments, PesTwin captures stochastic demographic effects, species-specific life-history traits, heterogeneous dispersal, and temporal variation in resource availability and infestation pressure. This enables mechanistic evaluation of outcomes that are difficult to resolve in generalisable compartmental models, including suppression dynamics, spatial spread, persistence, and resistance evolution.

We validate PesTwin against published laboratory cage data from four genetic control systems, drawn from three studies, in two insect species (\textit{Anopheles gambiae} and \textit{Drosophila melanogaster}), showing close agreement between predicted and observed population trajectories, including their replicate-to-replicate variability. We then illustrate how the same validated models extend beyond the cage to spatially explicit, field-scale scenarios, using PesTwin to explore how the timing, density and spatial placement of releases shape suppression and spread across heterogeneous landscapes. By making genetic control systems testable in silico before they are built or released, PesTwin can shorten the path from laboratory construct to field intervention: informing which constructs to prioritise, how to design the experiments that test them, where and when to release, and what evidence is needed to evaluate them.

% Please keep the Author Summary between 150 and 200 words
% Use first person. PLOS ONE authors please skip this step. 
% Author Summary not valid for PLOS ONE submissions.   
\section*{Author summary}
Genetic control technologies, engineered insects that spread traits through a population to suppress a pest or vector, or to block disease transmission, are advancing faster than the computational tools needed to predict their behaviour following release. Because almost none of these systems can yet be tested in the open field, their outcomes must be anticipated by simulation, yet existing models are typically either fast but coarse, tracking groups of identical individuals, or detailed but tied to a single species or question. We developed PesTwin, a flexible framework that represents every insect as an individual carrying an explicit genome and behaving within a realistic, spatially structured environment. PesTwin automatically derives the inheritance rules for complex engineered constructs, lets users assemble a model from reusable biological building blocks, and runs on Graphics Processing Units so that very large populations can be simulated. We show that PesTwin reproduces the detailed dynamics, including their natural variability, of four genetic control systems tested in laboratory cages, and we illustrate how the same models extend to predict how releases would unfold across a landscape. PesTwin is intended as a shared, reusable tool for the design, testing, and regulation of genetic control technologies, but its modular architecture extends naturally to non-genetic interventions such as sterile insect releases or Wolbachia-mediated cytoplasmic incompatibility.

\ifplossubmission
    \linenumbers
\fi

\section*{Introduction}
Vector-borne diseases and agricultural insect pests impose an enormous and persistent burden on human health and food security: vector-borne infections cause more than 700,000 deaths each year \cite{WorldHealthOrganization2024}, and insect pests destroy up to 40\% of global crop production \cite{Food2023} \cite{Savary2019}. Because this damage is concentrated in a relatively small number of species, it is in principle highly amenable to targeted intervention. In practice, however, the dominant tool, chemical insecticides, is losing ground: resistance is eroding efficacy across major pest and vector populations, regulation is narrowing the range of usable compounds, and broad-spectrum toxicity is increasingly recognised to harm non-target and beneficial species \cite{Wan2025}.

Biological control and habitat management can contribute, but their success is strongly context-dependent and has proven difficult to generalise across regions and target species \cite{Barratt2018}. There is, therefore, a pressing need for control strategies that are durable, species-specific, and scalable. 

Several complementary strategies meet this description to varying degrees, including the sterile insect technique (SIT), Wolbachia-based population suppression or replacement, and classical biological control using natural enemies; genetic control technologies represent the most recent and rapidly diversifying addition to this toolkit. Rather than repeatedly dosing the environment with chemicals, it introduces insects carrying engineered traits that spread through the target population, either reducing its size, typically through sterility or lethality, or rendering it unable to transmit disease \cite{Hcker2023}. 

The idea descends from the sterile insect technique, in which mass-reared sterile males are released to mate unproductively with wild females \cite{Dyck2005}. SIT is exquisitely species-specific and self-confining, but its dependence on sustained over-flooding (i.e., the excess of sterile males relative to those of the natural pest population) makes it logistically demanding. Programmable nucleases, and CRISPR in particular, have since produced a rapidly expanding repertoire of systems intended to overcome these limits: from self-sustaining gene drives that bias their own inheritance and spread autonomously  \cite{Gantz2015}, \cite{Hammond2016} to self-limiting designs such as split drives \cite{Kandul2020}, \cite{Terradas2021} and  Y-linked editors \cite{Burt2018}, \cite{Tolosana2025}], engineered to remain contained in space and time, reviewed in \cite{Han2024}. Crucially, these systems are not interchangeable: they differ profoundly in how strongly and how durably they suppress a population, how far they spread, how readily they can be reversed, and how vulnerable they are to the evolution of resistance.

This proliferation of designs has created a challenge for evaluation. Almost none of these systems can yet be tested in the open field, so their behaviour must be anticipated computationally, and the questions most relevant to deployment are precisely those that simple models cannot resolve. Whether a release succeeds depends on the interplay of stochastic demography, spatial structure, species-specific life history, and the detailed mechanics of inheritance, and on operational choices such as when, where, how densely, and how often to release, how to combine genetic control with conventional methods, and how to manage resistance. A framework capable of informing such decisions must therefore couple stochastic, individual-level population dynamics with explicit space and with a mechanistic but scalable representation of inheritance, and it must remain flexible enough to place very different systems on a common footing.

Existing tools meet only part of this specification, and they do so in instructive ways. The most widely used, MGDrivE \cite{SnchezC2020}, since extended to incorporate seasonality and epidemiology \cite{Wu2021}, \cite{Mondal2024}, achieves efficiency and generality by tracking the number of individuals in each genotype class across a network of patches; the price is that it does not resolve fine-scale space or individual-level agent-to-agent interactions within a patch, and its inheritance rules must be written by hand, which becomes intractable as constructs grow more complex. Individual-based models recover this lost detail by simulating discrete organisms, but the available frameworks were each built for a particular purpose. EMOD \cite{Bershteyn2018}, \cite{Eckhoff2017}, \cite{Selvaraj2020} and Skeeter Buster \cite{Magori2009} resolve mosquito ecology and gene-drive dynamics in rich detail, yet remain tied to specific vector–host or single-species systems, while SLiM \cite{Messer2013}, \cite{Haller2017}, \cite{Haller2019}, \cite{Haller2023}, \cite{Haller2026} is a powerful engine for long-term evolution rather than for short-term, spatially explicit operational forecasting. Beyond these, much of the field relies on bespoke models built for individual constructs (e.g., \cite{North2020}, \cite{Hancock2024}, \cite{Beaghton2019}), which are hard to compare or reuse. What remains missing is a single framework that unites stochastic population dynamics, flexible spatial structure, and a mechanistic yet scalable genetic representation, and that can be reused across species and across genetic designs.

\begin{figure}[!htb]
    \centering
    \includegraphics[width=1\linewidth]{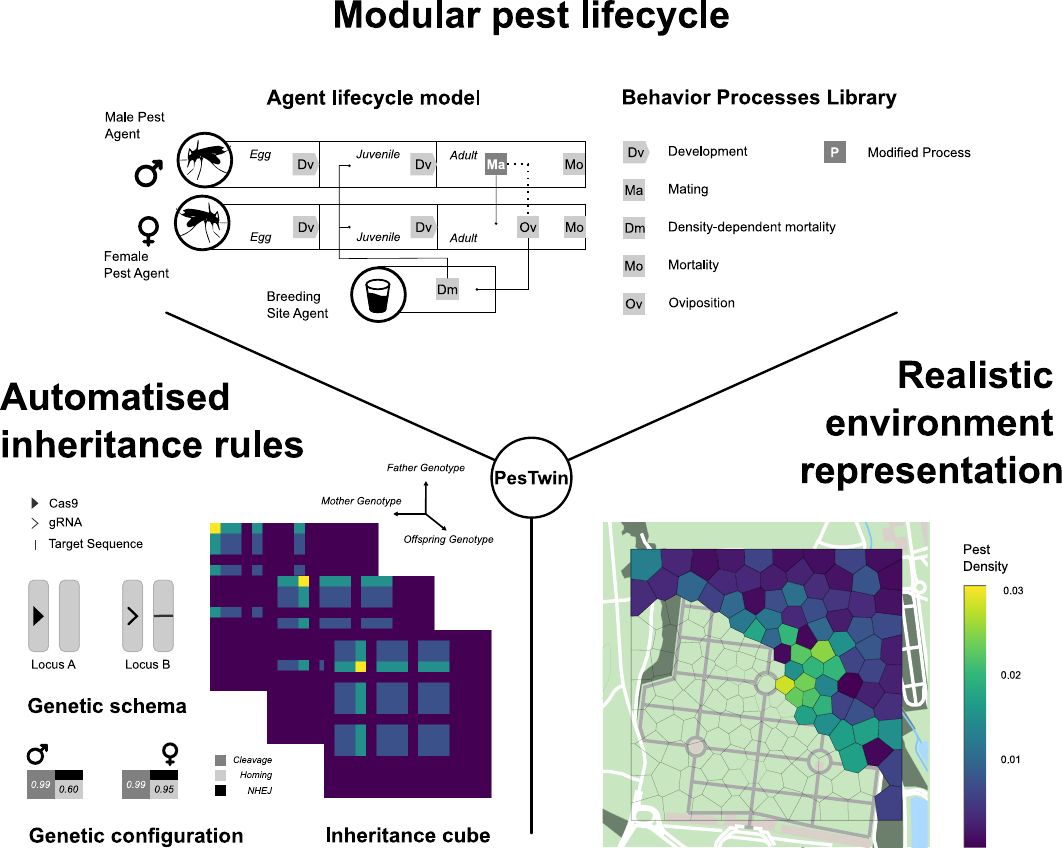}
    \caption{Graphical abstract of the PesTwin modelling and simulation framework. A modular pest-lifecycle module allows agent-based models of increasing complexity to be assembled from predefined behavioural processes; the parameters of these processes can be modified according to the genotype carried by each agent, providing a direct mapping from genetic state to phenotype. An Inheritance Cube Generator automates the construction of inheritance rules, so that complex engineered systems, built from multiple transgenic constructs distributed across several loci, can be modelled without enumerating genotype combinations by hand. Space is represented flexibly as a directed graph that natively incorporates GIS coordinates, narrowing the gap between simulation and realistic environmental conditions.
    }
    \label{fig:graphical_abstract}
\end{figure}

Here we present PesTwin, an agent-based framework designed to fill this gap. Every insect is modelled as an individual agent that carries an explicit genetic state and behaves within a graph-structured, spatially explicit environment, so that demographic stochasticity, heterogeneous dispersal, and species-specific life history emerge naturally from the simulation. Three components give the framework its flexibility (Fig. \ref{fig:graphical_abstract}): an automated Inheritance Cube Generator that builds mechanistic inheritance rules for arbitrary multi-locus constructs; a modular lifecycle in which behavioural processes can be assembled and reconfigured without rewriting the model; and a graph-based spatial layer representing heterogeneous landscapes and the dispersal pathways between them. Together they allow genetic control systems of very different design to be encoded, validated against data, and compared within one environment. In what follows, we describe the framework, validate it against published cage-trial data from four CRISPR-based systems in \textit{Anopheles gambiae} and \textit{Drosophila melanogaster}, and show how the same models extend to spatially explicit, field-scale scenarios.

\section*{Design and implementation}
\subsection*{Modelling scope and architecture}

To represent both natural pest population dynamics and their response to  control, we identified five modelling domains that determine whether population-level outcomes can be predicted with operational realism, designed for flexibility across species and environments: 
\begin{itemize}
    \item \textbf{Life history and development} - species-specific developmental stages, reproduction, survival, and behaviour across diverse arthropod pests and vectors.
    \item \textbf{Genomic architecture and inheritance} - chromosomal composition and inheritance, recombination, sex determination systems, fitness effects, and engineered genetic control mechanisms.
    \item \textbf{Host interactions} - interactions between pests and crops, animal hosts, or intermediate hosts to quantify agricultural damage or disease transmission.
    \item \textbf{Local and regional dispersal} - local dispersal, long-range migration, and environmental drivers of movement that influence outbreak dynamics and intervention spread.
    \item \textbf{Environmental heterogeneity} - spatial and temporal variation in breeding, feeding, and resting habitats, including seasonal, climatic, and agricultural effects on population dynamics.
\end{itemize}

These domains are deliberately general: while our validation focuses on genetic control systems, the same domains and process library apply to other control strategies, such as SIT, Wolbachia, or even chemical pesticides, which act primarily through the Life history/development and Host interactions domains rather than through Genomic architecture and inheritance, as previously demonstrated for a SIT intervention against \textit{Drosophila suzukii} in our earlier work in a precision-agriculture setting \cite{DeAntoni2026}.

While these domains determine population-level outcomes, they operate across multiple biological scales, from individual-level genetic inheritance and development to population- and landscape-scale processes such as dispersal, host interaction, and environmental heterogeneity, requiring explicit representation of interacting agents and their environment. A well-known limitation of agent-based modelling frameworks is scalability and computational performance. Computational cost scales with agent count, and without mitigation, performance degrades rapidly as population sizes grow. GPU-accelerated parallel computing addresses this bottleneck directly, and PesTwin exploits this by building on FLAMEGPU \cite{PRichmond2021}, an open-source library that executes agent-based simulations on GPU hardware and thereby enables population sizes that would be prohibitive on conventional processors. A preliminary assessment demonstrated that the current software can sustain simulations of tens of millions of concurrent agents, equivalent to simulating a field of tens of hectares, depending on species and infestation density. This scale is appropriate for the intended application: a Decision Support System for monitoring and managing an individual field. All simulations presented in this work were run on an NVIDIA A10G GPU (24 GB VRAM), using FLAMEGPU version 2.0.0rc1.

\subsection*{Automated inheritance: the Inheritance Cube Generator}
To simplify configuration of diverse genetic control systems, we developed the Inheritance Cube Generator, a Python package that automates construction of inheritance rules for mechanisms including CRISPR-Cas9, toxin–antidote, and RNA interference systems. Our approach builds on the concept of an inheritance cube for representing genetic transmission \cite{SnchezC2020}, where inheritance is encoded as a three-dimensional tensor of dimensions $(N, N, N)$, with $N$ the number of genotype classes. For each parental genotype pair $(i, j)$, indexing the cube at position $[i, j]$ returns the full distribution of offspring genotypes.
In PesTwin, this representation is generated directly from the genetic structure of the system using a modular design based on alleles, loci, and chromosomes. Alleles are defined by locus, loci are organised into chromosomes, and chromosomes into genomes, allowing multi-locus systems (including engineered genetic control elements) to be represented without specifying all genotype combinations and outcomes. For CRISPR-based systems, gRNA-driven Cas9 induces cleavage at target sites when both molecules are expressed, with repair resolved through competing pathways: homology-directed repair (HDR), which underlies homing-based systems including self-sustaining gene drives and self-limiting derivatives, and end-joining (EJ), which generates mutational outcomes that can produce resistance to Cas9 cleavage but also forms the basis of suppression strategies such as Y-linked editors and X-shredders. Where and when these mutations occur may also affect the fitness of the progeny, as well as the rate and type of DNA-cleavage repair outcomes \cite{Hammond2021}.

Within this representation, inheritance is implemented through the generated cube, which maps parental genotypes to offspring distributions. The Inheritance Cube Generator is provided as a standalone package and can be integrated into other modelling frameworks. It includes:
\begin{itemize}
    \item \textbf{Sex determination systems}: since certain alleles linked to sex (e.g., Y in an XY system, or M in M/m systems) constrain the number of viable genotypes, combinations such as YY and invalid mating pairs are excluded.
    \item \textbf{Chromosomal recombination}: input on recombination between linked loci during meiosis at pre-defined frequencies, enabling reshuffling of genetic elements into new offspring combinations. 
    \item \textbf{Germline modification}: genome editing occurs in the germline of individuals expressing the construct(s), biasing inheritance ratios, so transmitted genotypes can differ from the parental genotype.
    \item \textbf{Embryo modification}: genome editing occurs in the embryo post-fertilisation, driven by parental deposition of nuclease components and independent of construct inheritance, leading to genotypic and phenotypic changes that depend upon the parental genotype and are typically biased towards EJ over HDR. 
\end{itemize}

Thus, PesTwin removes the need to specify parental genotype combinations and offspring outcomes manually, which becomes rapidly intractable and error-prone as system complexity increases. A detailed description of the Inheritance Cube Generator algorithm can be found in \nameref{si_inheritance}.

\subsection*{Modular agent lifecycle}
In an agent-based model, each agent (the insect) is represented as an autonomous entity that interacts with other agents (insects) and the environment through a set of behavioural rules. In PesTwin, these rules are implemented as modular processes that act on an insect’s internal state, the set of variables that define its current biological condition. This includes its life stage, location, reproductive status, survival status, and genetic identity. The environment is similarly rich, representing breeding sites, food sources, and hosts. At each simulation step, processes read and update the state variables, producing outcomes such as movement, mortality, reproduction, or life-stage transition. By including context-dependent behavioural rules and a graph-like spatial representation, PesTwin can reproduce complex system behaviours such as density-dependent larval mortality and dispersal.

The flexible nature of PesTwin allows users to build models that can be adapted to different species, environments, and research questions. This flexibility is achieved through a modular design, where behaviours are implemented as reusable and switchable components that can be assigned to different agent states and thus combined in various ways. Users with varying backgrounds and expertise can define simple rules by linking behaviours together; for example, the event of egg-laying only occurring after a successful mating event, thus making it possible to create models of increasing complexity without redesigning them from scratch. The main processes represented in the model can be grouped into a) life cycle processes such as development and mortality (insect does or does not progress to the following life-stage, respectively), b) habitat use processes, such as dispersal (distance of movement), breeding-site identification and selection, and density-dependent mortality (extra mortality counts driven by scarce resources or competition), and c) reproduction processes, such as mating (selection of partner genotype) and offspring production (generation of new agents). PesTwin is therefore intended to be used across the genetic-control pipeline rather than by modellers alone: molecular biologists and technology developers specify the genetic construct and its expected molecular behaviour; modellers encode this within the framework, parameterise the relevant life-history and environmental processes, and run the simulations; and the resulting population-level predictions, suppression, spread, persistence, and resistance, are reported in terms that experimentalists, program managers, and regulators can interpret directly. In this way, the same model can act as a shared language linking these groups. Moreover, the approach is designed to produce visual outputs that facilitate communication of expected outcomes and uncertainty to stakeholder groups of varying familiarity with genetic control.

In the ideal case, configuring a new model requires no programming: the user selects an appropriate set of processes from the existing library, specifies their sequencing logic, for instance, conditioning oviposition on a prior successful mating event, and calibrates the resulting model using species-specific empirical data. Introducing a genuinely new behaviour, for example, blood-feeding in mosquitoes, would instead require implementing a new process, but the object-oriented design makes this a localised addition rather than a redesign of the model. The modelling of genetic control technologies is made possible by using process modifiers. A modifier associates the agent’s genetic information with a specific success rate, a number between 0 and 1. For example, a transgene known to induce lethality at the pupal stage can be implemented as a multiplier of the larva-to-pupa stage transition, setting its success rate to zero. In such a case, the agents carrying the construct will never reach the pupal stage, while agents not carrying such a construct will not be affected. In addition to their own genotype, agents can be assigned other pieces of genetic information, such as referencing the genotypes of their parents. This information can be exploited by using a parental modifier to model phenomena like reduced fitness or cleavage due to CRISPR parental deposition. From a software engineering perspective, agent processes are implemented using object-oriented principles, with extensive use of class inheritance to provide standard interfaces, promote code reuse, and simplify the development of new behaviours. Additional implementation details are provided in \nameref{si_agents}, following the Overview, Design, and Details protocol for agent-based model reporting \cite{Grimm2006}, \cite{Grimm2010}.

\subsection*{Spatially explicit environment}
In PesTwin, the spatial environment is modelled as a directed graph whose nodes correspond to discrete locations, each optionally associated with GIS coordinates. Agent interactions are strictly local: two agents can only interact if they occupy the same node at the same simulation step. An edge between two nodes exists whenever there is a nonzero probability of movement between the corresponding locations. Dispersal is decomposed into two independently configurable components: an \textit{exit} function, governing whether an agent departs its current node at each time step, and a \textit{move} function, determining which neighbouring node it selects upon departure. This modularity allows a wide range of movement behaviours to be specified without restructuring the underlying graph. For example, simple diffusion can be represented by combining a constant exit probability with a random selection among available destinations, represented by node edges. More realistic movement patterns can be obtained by assigning probabilities to edges based on factors such as the distance between locations, or the population density and resource availability at the source and destination nodes, and by constraining agents with a maximum daily travel range. Because dispersal is expressed as agent movement on this graph, the same machinery extends naturally to passive, host-mediated transport: a mobile host can be represented as an additional agent whose movement carries the non-dispersing life stages associated with it. This makes it possible to capture phenomena such as the spread of New World screwworm (\textit{Cochliomyia hominivorax}), whose eggs and larvae are transported between sites on infested vertebrate hosts such as deer or cattle. In the applications presented in the following section, this framework is used to capture both short-range daily movements and longer-range migration over extended periods. Parameters for dispersal models can be estimated from mark–release–recapture studies \cite{Marini2010}, \cite{Marini2019} or from more advanced close-kin mark–recapture approaches \cite{Sharma2022}, \cite{Marshall2025}.

One of the advantages in adopting an agent-based approach is its flexibility in capturing the nuances of dispersal, which are often hard to capture in other modelling frameworks such as compartmental or patch-based models. As an example in PesTwin, the exit function implementation could easily link the probability of leaving an environmental node to the abundance of a specific resource at a given location, encouraging agents to move when conditions such as food scarcity arise. Similarly, the move function can represent different movement mechanisms, ranging from autonomous navigation, such as following a gradient of sensory clues, to environmental dispersal, such as transport by wind or responses to altitude gradients.

% Results and Discussion can be combined.
\section*{Results}
To validate PesTwin and demonstrate its capabilities, we reproduced three published cage-trial studies of CRISPR-based genetic control, spanning two insect species and four genetic systems \cite{Strampelli2025}, \cite{Hammond2021}, \cite{Terradas2021}. Each study was chosen to demonstrate a different combination of framework features, homing inheritance, resistance evolution, parental deposition, split genetic architecture, and contrasting sex-determination and life-history configurations, and to test whether PesTwin reproduces not only the mean population trajectory but also the replicate-to-replicate variability observed experimentally. For every system, the relevant life-history and genetic processes were parameterised from the original study, and replicate simulations were compared directly against the published cage data. Accurate description of models and parameters can be found in \nameref{si_models}.

In every panel, we deliberately restricted the comparison to what the original experiments could measure, typically the frequency of a transgenic phenotype (for example, whether individuals carry one or more of the engineered constructs), and generally without resolving the underlying genotype or zygosity. Overlaying the simulations on this directly observable quantity grounds the validation in what was actually recorded. The model itself, however, resolves a great deal more: the full distribution of genotypes, the individual and population-level fertility, and the rise and fall of distinct classes of resistance allele, among other outputs. Beyond validating the model, this richer information is operationally valuable: the genotype and allele-frequency dynamics underlying an observed phenotype can guide the detection and monitoring of resistance, through genotyping of individuals, pooled amplicon sequencing of bulk samples, or diagnostic combinations of phenotypic markers such as reduced fertility or incomplete sexual development, supporting earlier detection of, and response to, resistance and other emergent outcomes in both field and laboratory settings.

\begin{figure}[!htb]
    \centering
    \includegraphics[width=1\linewidth]{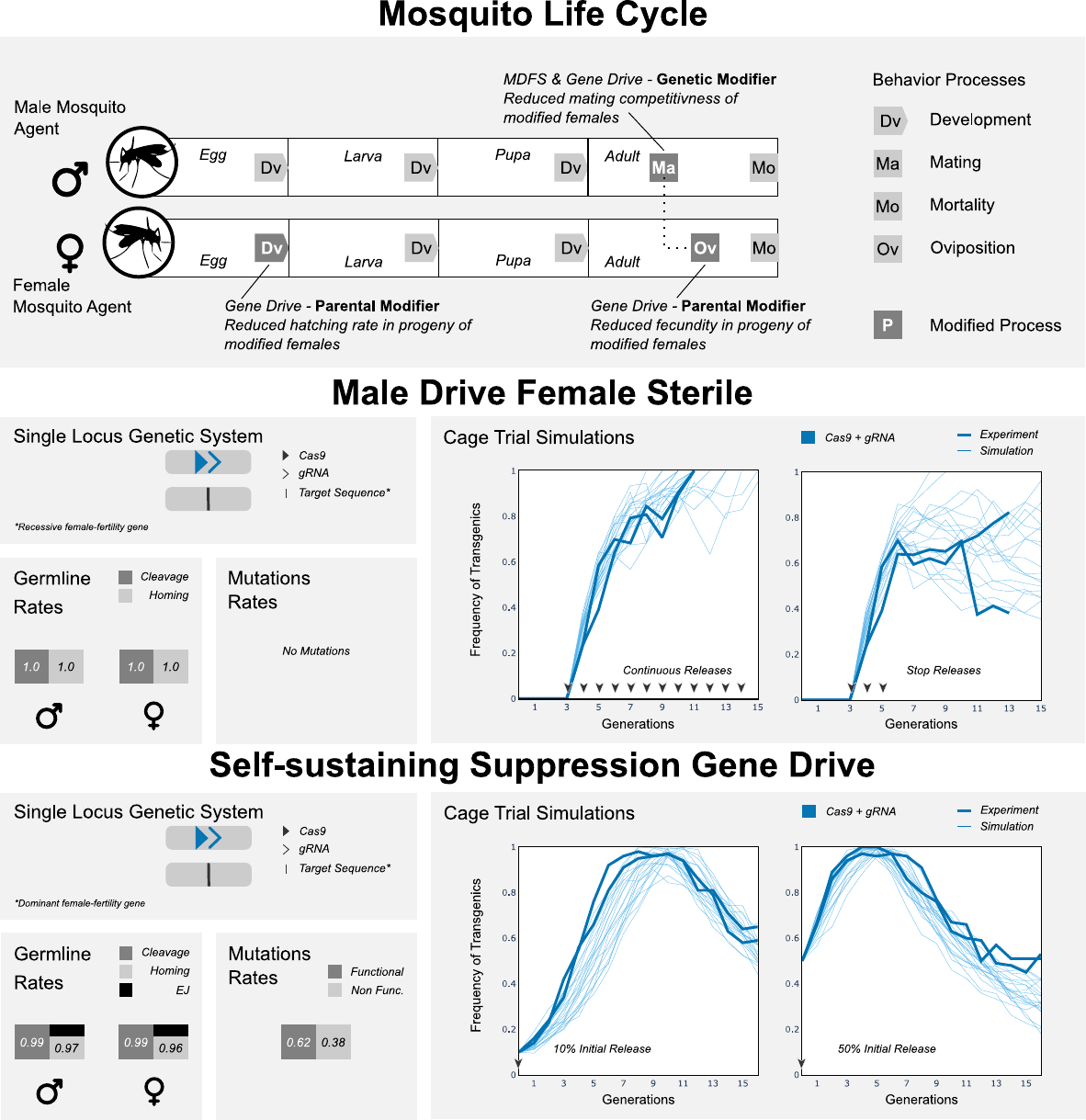}
    \caption{PesTwin simulations recapitulate cage trial dynamics for two homing-based genetic control systems in \textit{Anopheles gambiae}. In all population panels, bold lines show the experimental cage data and thin lines the overlaid PesTwin simulations. \textbf{(A)} Schematic of the agent-based \textit{A. gambiae} lifecycle, in which genotype-specific modifiers act on selected processes: here adult male mating success, oviposition success, and egg-to-larva development through a reduced hatching rate. \textbf{(B)} Male Drive Female Sterile (MDFS). The genetic model places Cas9 and the gRNA together at a single locus, opposite the target site at the homologous position on the partner chromosome, and resolves germline cleavage and mutation rates (a single mutational class, with no embryo cleavage). Two cage experiments are shown, each with two replicate releases and overlaid with 20 PesTwin simulations: continuous release of MDFS males from generation 3 onwards, driving population eradication (left), and a release limited to generations 3–5, after which the population recovers (right). \textbf{(C)} Suppression gene drive. The genetic model shares the single-locus architecture of MDFS but additionally resolves, among cleaved chromosomes, the fraction of events repaired by end-joining (EJ) versus homing, specified separately for the germline and the embryo, and the fraction of EJ outcomes that yield functional (r1) versus non-functional (r2) resistance alleles. In this system, the target gene disrupted by the gene drive is a haplosufficient female fertility gene, leading to female sterility in females carrying the gene drive in homozygosity. Two cage experiments are shown, seeded at 10\% (left) and 50\% (right) initial transgenic frequency from generation 0, and each overlaid with 20 PesTwin simulations; in both, transgenic frequency rises to a peak before declining as cleavage-resistant alleles emerge and are selected.}
    \label{fig:results_cage_mosquitoes}
\end{figure}

\subsubsection*{Male drive female sterile}
The first system is a Male Drive Female Sterile (MDFS) construct, a self-limiting suppression approach developed for the malaria vector \textit{Anopheles gambiae} \cite{Strampelli2025}. A complete CRISPR cassette carrying both Cas9 and a gRNA is inserted into a gene required for female fertility, so that females carrying the construct are sterile, whereas males suffer no measurable fitness or fertility cost. Because homing occurs in the male germline, the construct is inherited at super-Mendelian rates, exceeding the 50\% expected under Mendelian inheritance, and spreads rapidly through successive generations. The strategy is self-limiting: as inheritance is biased only through males, the construct cannot be maintained indefinitely, and suppression is therefore bounded in time and space.

In PesTwin, this system exercises three core features of the framework: homing implemented through the inheritance cube, a genetic modifier encoding genotype-dependent female sterility, and parameterisation of individual life-history processes from laboratory data. The model used here is deliberately simplified relative to the original study in that it omits CRISPR-resistant alleles, yet it reproduces the key dynamics of the system (Fig. \ref{fig:results_cage_mosquitoes}): the spread of the transgene over several generations, the collapse of the caged population under repeated releases of transgenic males, and the self-limiting recovery of the population, accompanied by a decline in transgene prevalence, once releases are stopped.

Beyond these mean trends, the ensemble of PesTwin simulations reproduces the characteristic divergence among trajectories that emerges late in suppression and after releases cease, matching the variation seen across the four replicate cages of the original experiment. This behaviour reflects a transition from deterministic to stochasticity-dominated dynamics: during the release period, repeated introductions act as a strong deterministic forcing that keeps replicates tightly clustered, but as the construct approaches fixation, or once releases end, this forcing weakens. Under continuous release, the residual variability is driven largely by genetic drift at diminishing wild-type population sizes; under pulsed release, it additionally reflects the opposing pressures of negative selection against sterile females and the positive invasive pressure of homing, whose balance becomes increasingly unpredictable at small population sizes. The agreement between the dispersion among simulated trajectories and the observed replicate variation indicates that PesTwin captures not only the mean behaviour of MDFS invasion but also its inherent variability, a property that is difficult to resolve in deterministic, population-level models.

\subsubsection*{Self-sustaining suppression gene drive}
The second system is a self-sustaining suppression gene drive, again developed in \textit{Anopheles gambiae} \cite{Hammond2021}. Like MDFS, it suppresses the population by targeting a gene required for female fertility, but it is designed to persist and spread rather than decay. Because a single copy does not fully sterilise females, heterozygous females remain fertile and continue to transmit the drive; the construct also copies itself onto the homologous chromosome (a process known as homing) and therefore continues to invade the population even as numbers decline. The study reproduced here focuses on how the choice of target site governs the evolution of resistance, because imperfect homing can be repaired by end-joining to create functional, drive-resistant alleles.

Modelling this system in PesTwin requires representing co-occurring homing and end-joining repair in the germline, together with the fitness consequences of CRISPR-component deposition in the embryo by transgenic parents; the latter are implemented through parental modifiers that adjust offspring life-history according to parental genotype. Following a single release of transgenic males, the simulated drive spreads rapidly and almost to fixation, driving a steep decline in population size (Fig. \ref{fig:results_cage_mosquitoes}). As the population bottlenecks, however, functional resistant alleles gain a selective advantage and accumulate, progressively reducing drive efficiency, halting further spread, and ultimately curtailing long-term suppression.

Here, PesTwin was used to generate a large ensemble of stochastic simulations, which we compared against the four replicate cages reported in the original study \cite{Hammond2021} (Fig. \ref{fig:results_cage_mosquitoes}). The simulations closely reproduce the observed dynamics. In contrast to MDFS, the drive shows moderate stochasticity from shortly after release that does not grow substantially over time, and peak drive frequency is predicted at generations 8–12 for the 10\% release and generations 4–6 for the 50\%, in agreement with the data. The model also captures the timing and consequences of resistance: consistent with earlier theory \cite{Beaghton2019}, \cite{Hammond2017}, selection for resistance is weak early in the spread, when the drive is largely heterozygous and female sterility is low, but it intensifies as the drive nears fixation and the few remaining fertile females are disproportionately those carrying resistant alleles, producing a rapid reversal from spread to loss. This reversal does not proceed at a constant rate: as drive frequency falls and resistance rises, selection against the drive weakens, slowing its elimination and leaving drive and resistance frequencies to plateau rather than the drive being cleanly removed, behaviour that PesTwin recovers across the simulated ensemble.

\subsubsection*{Split gene drives}
The third case comprises two \textit{Drosophila melanogaster} split-drive systems \cite{Terradas2021}, in which the Cas9 and gRNA components are placed at separate loci and inherited independently; the gRNA can home only when Cas9 is also present. The two variants differ in the genomic location of Cas9 (autosomal or X-linked). Unlike the previous cases, the objective here is population replacement rather than suppression: although the target sites are essential genes, a recoded second copy is supplied so that CRISPR activity does not generate lethal loss-of-function genotypes.

\begin{figure}[!htb]
    \centering
    \includegraphics[width=1\linewidth]{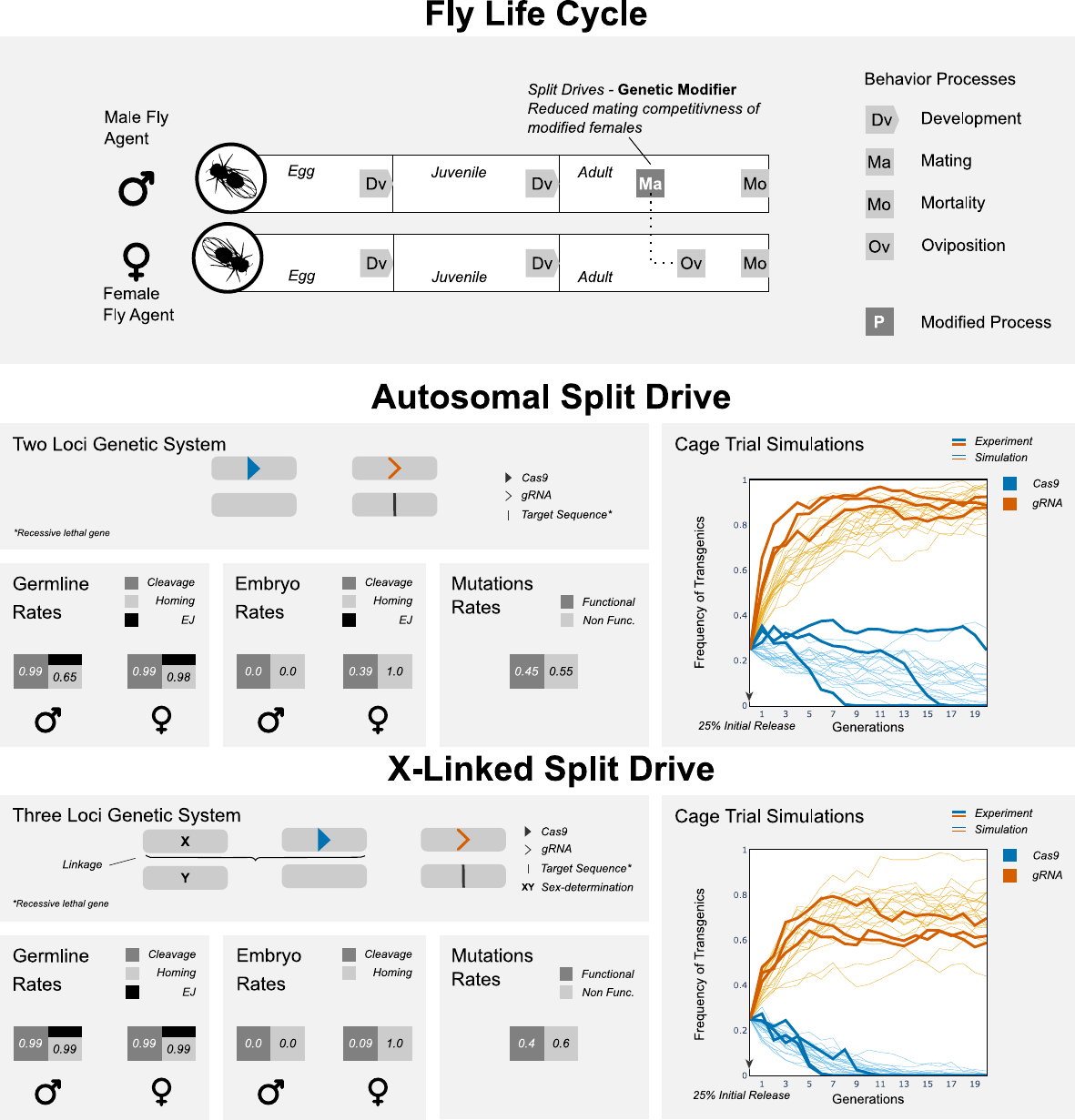}
    \caption{PesTwin simulations recapitulate cage-trial dynamics for two split-drive systems in \textit{Drosophila melanogaster}. In all population panels, bold lines show the experimental cage data and thin lines the overlaid PesTwin simulations. \textbf{(A)} Schematic of the agent-based \textit{D. melanogaster} lifecycle: egg, juvenile and adult stages, with sex-specific mating, oviposition and mortality, illustrating how a genotype-specific modifier can alter a life-history trait; reduced mating competitiveness is shown here as an illustrative example. \textbf{(B)} Autosomal split drive. The genetic model places Cas9 and the gRNA on two separate chromosomes, with the target gene at the gRNA’s homologous insertion site, and resolves cleavage rates and, among cleaved chromosomes, the fraction repaired by end-joining (EJ) versus homing, specified separately for germline and embryo, and the fraction of EJ outcomes yielding functional (r1) versus non-functional (r2) resistance alleles. The frequencies of individuals carrying Cas9 and/or the gRNA are tracked over time across three replicate cages and overlaid with 20 PesTwin simulations. \textbf{(C)} X-linked split drive, with Cas9 linked to the X chromosome and the gRNA and target gene on a separate chromosome; model components are as in \textbf{(B)}, tracked across three replicate cages and overlaid with 20 PesTwin simulations. In both configurations the gRNA drives through the population before stabilising, as the Cas9 element is progressively lost and cleavage-resistant mutations accumulate.
    }
    \label{fig:results_cage_flies}
\end{figure}

This case exercises further features of the framework: explicit representation of the XY sex-determination system, incorporation of embryonic Cas9 deposition into the inheritance cube, and flexibility in lifecycle design, only three stages (egg, a combined juvenile stage, and adult) are modelled, because the cage data track discrete, non-overlapping generations scored as adults, so immature stages that do not affect the genotype dynamics can be collapsed without loss of accuracy. We simulate the trajectories of both transgenic components and compare them against the cage data (Fig. \ref{fig:results_cage_flies}).

The split-drive architecture, two independently segregating elements acting in concert, together with two classes of resistance allele, produces more stochastic invasion dynamics than MDFS, and PesTwin captures this behaviour for both the autosomal and X-linked configurations (Fig. \ref{fig:results_cage_flies}). The gRNA, which homes when co-inherited with Cas9, initially spreads with low variance and high reproducibility, echoing the early deterministic phase of MDFS, before substantial stochasticity emerges within a few generations, a transition the model reproduces. Agreement is particularly close for the X-linked configuration; for the autosomal configuration, one replicate spread slightly faster than predicted even allowing for the stochasticity of the model, which may reflect a statistical outlier, slightly insufficient stochasticity in the parameterisation, or a founder effect in which post-release mating bias seeded a stronger initial invasion.

The Cas9 element behaved very differently: rather than spreading, it was progressively lost, and markedly faster in the X-linked than the autosomal configuration. This follows from the partially essential nature of the target gene, \textit{prosalpha2}: null genotypes are strongly counterselected, and, as described in the original study \cite{Terradas2021}, sub-Mendelian transmission of Cas9 is observed in experimental crosses, consistent with the reduced viability of individuals carrying both components. Consequently, although the gRNA homes and spreads, it too experiences counterselection through mutation of \textit{prosalpha2}, and individuals carrying both elements are at a fitness disadvantage. PesTwin captures these opposing pressures, the invasive capacity of the gRNA balanced against selection imposed by the target gene, and reproduces the observed pattern in which the gRNA persists and spreads while Cas9 is gradually eliminated from the population.

\subsection*{Simulating environmental release}
Genetic control technologies, from self-limiting releases to self-sustaining gene drives, remain largely confined to contained laboratory and cage studies, and any move toward an open release will rest heavily on what can be anticipated about its consequences \cite{GeneDrives2016}. Because those consequences cannot yet be observed in the field, they must be explored computationally. Modelling is valuable here not only for estimating how effective an intervention is likely to be, but also for weighing its cost and operational burden and for assessing the risks it may carry. These questions, moreover, are not fixed: the relevant operational setting (e.g., a farmer's field versus an urban neighbourhood), the spatial scale, and even the outcome of interest, whether the degree of suppression, the evolution of resistance, or the persistence of an engineered trait, vary from one deployment to another, and a useful framework must be able to accommodate all of them.

These considerations apply across the whole spectrum of genetic control, but they carry particular weight for self-sustaining systems. Self-limiting strategies are bounded in time and space and can, in principle, be discontinued, whereas gene drives are designed to spread through a population across space and time, can have large and durable effects, and cannot easily be recalled once released; the consequences of a release are correspondingly greater, and the ability to anticipate outcomes before a decision is made all the more important. In this lies the value of digital tools such as PesTwin: they allow the research and operational questions for which real-world data are still missing to be investigated in advance, and, as we show below, explored under realistic, spatially explicit conditions.

\begin{figure}[!htb]
    \centering
    \includegraphics[width=1\linewidth]{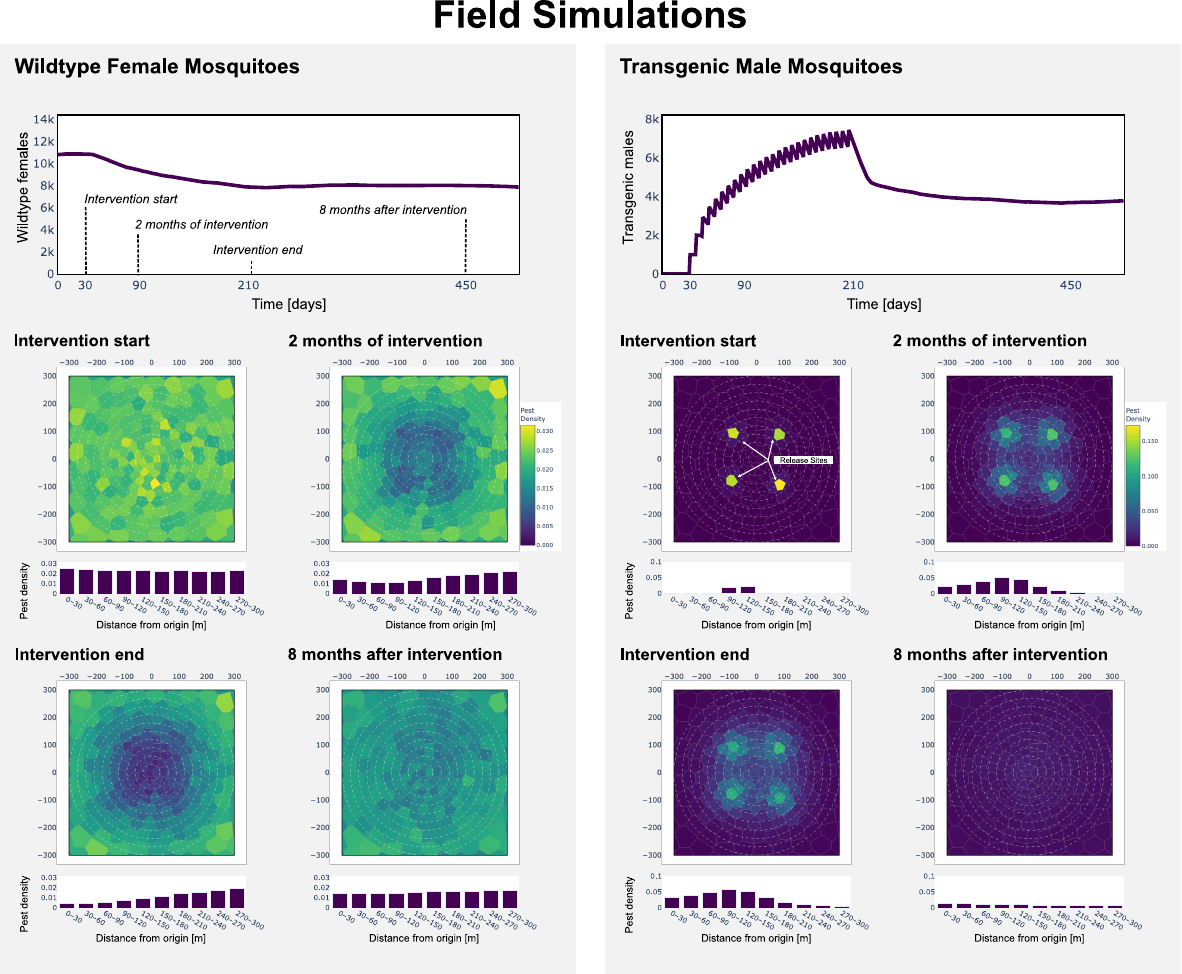}
    \caption{Simulation of a six-month pest-control intervention through weekly releases of MDFS males in a spatial landscape. The site is about 49 hectares in extent, and transgenic males are introduced at four release sites equidistant from the origin; the mean of ten stochastic simulations is shown. \textbf{Left panel}: population dynamics of wildtype female mosquitoes during and after the intervention. The top plot shows the overall population trace declining from its initial equilibrium as a result of the intervention. The spatial distribution of pest density is shown over a 300 m radius area at four time points, at the start, during, at the end, and eight months after the intervention, with the lower inset plots showing pest density as a function of distance from the origin; the four release sites shape a zone in which the target population is effectively suppressed, and once releases stop the diffusion of transgenic insects across the landscape dilutes the suppression over the whole area. \textbf{Right panel}: population dynamics of the released transgenic males. The top plot shows the transgenic male population accumulating during the intervention and then stabilising once releases are interrupted, and the spatial snapshots show the diffusion of transgenic males around the release sites, which creates the origin-centred zone of suppression for the target female population.
    }
    \label{fig:field_results}
\end{figure}

To demonstrate the capabilities of the framework in modelling insect spatial dispersal, we simulated an idealised scenario of pest-control intervention; the results are shown in Fig. \ref{fig:field_results}. The scenario consists of a square landscape of about 49 hectares (roughly 700 m on a side), inhabited by \textit{Aedes aegypti} mosquitoes. The simulated control is based on weekly releases of MDFS transgenic males over six months, after which the site is monitored for a further ten months. Four release sites equidistant from the origin (100 m) are used, and at each site 250 males are released, for a total of 1,000 transgenic males introduced across the area each week. Space is represented as a non-uniform tessellation of the intervention area, with cells whose radius increases with distance from the origin; a breeding-site agent is instantiated in each cell, corresponding to a node of the dispersal graph. The capacity of each breeding site is assumed to scale with the area of its cell, so that a homogeneous distribution of pests is expected at equilibrium. Carrying capacities were chosen to give an adult population density of a few hundred females per hectare, consistent with published estimates in the literature \cite{Carvalho2015}, \cite{Villela2015}, \cite{Ritchie2013}.

As transgenic males are released, they mate with wildtype females and generate transgenic offspring, in particular females that are unable to bite or mate. Both released and newly emerged transgenic males disperse across the graph. During the intervention, the presence of transgenic males around the release sites sustains a localised zone around the origin in which the target wildtype female population is effectively suppressed, with density falling to less than 20\% of its initial value. This suppression is local and weakens with distance from the origin and the release sites. Once the intervention ends and no further transgenic males are introduced, the transgenic population stabilises and slowly diffuses across the whole area, producing a new wildtype-female equilibrium that remains suppressed relative to the initial state. This persistence of transgenics after releases stopped is a characteristic of the MDFS design, arising specifically because males carrying the construct bear no direct fitness cost.

\section*{Conclusions}
The evaluation of genetic control technologies is constrained by the absence of field-scale deployments, meaning that inference must be drawn primarily from cage experiments that simplify spatial structure and environmental variability. While these systems capture key genetic processes, they do not reflect the full chain of effects that emerges in real settings, where molecular events at the level of DNA inheritance translate through population dynamics into spatially distributed ecological outcomes. This created a persistent discontinuity between mechanistic genetic design, laboratory population behaviour, and the conditions under which control strategies would be deployed.

PesTwin is designed to bridge this discontinuity by linking three levels of representation within a single framework: mechanistic genetic modelling that can represent complex inheritance and molecular processes, stochastic population dynamics consistent with well-mixed cage-like systems, and explicit spatial environments that approximate field conditions. This structure allows the same underlying model to reproduce both controlled experimental outcomes and emergent behaviours that only arise when populations interact across heterogeneous landscapes. Across three validation cases, the framework recapitulates known cage dynamics, while also capturing higher-order effects such as temporary population increases under suboptimal control, for example when reduced larval density lowers competition and increases survival, see \nameref{si_overcompensation}.

More broadly, this integration provides a basis for translating genetic constructs into operational insight, supporting the design of release strategies, monitoring schemes, and risk-relevant outcome measures for disease and pest control. It also enables explicit evaluation of biosafety-relevant behaviours, including spatial spread and persistence under different intervention regimes, while providing a consistent computational language through which predictions can be communicated to stakeholders across scientific and regulatory contexts. 

Although the systems validated here are CRISPR-based homing and split drives, PesTwin is not tied to any single mode of action. Because genetic behaviour enters the model through the Inheritance Cube Generator and through genotype-specific process modifiers, the same framework can represent a broad range of technologies, including repressible female-lethal systems based on tet-off elements, X-shredders driven by endonucleases such as I-PpoI, and toxin–antidote systems, among others. More broadly, because the modular life cycle and process-modifier design do not require a genetic mechanism specifically, the same framework extends naturally to non-genetic control strategies: SIT, Wolbachia-based, or canonical biocontrol interventions which act on the same demographic and behavioural processes. The modular design likewise allows differences between laboratory and field conditions to be incorporated through the spatial layer and environment-dependent processes as data accumulate, rather than being fixed in advance. This generality, together with the framework’s reusable architecture, is intended to let PesTwin keep pace with a rapidly diversifying field and to serve as a common platform on which very different control strategies can be compared.

Overall, PesTwin provides a unified framework linking genetic mechanism, laboratory population dynamics, and field-scale ecological outcomes, with the aim of enabling more reliable prediction of efficacy, behaviour and biosafety as these technologies move toward deployment.

\section*{Supporting information}

% Include only the SI item label in the paragraph heading. Use the \nameref{label} command to cite SI items in the text.

\paragraph*{S1 Appendix.}
\label{si_inheritance}
{\bf Inheritance cube generator.}

\paragraph*{S2 Appendix.}
\label{si_agents}
{\bf Agent life processes.}  

\paragraph*{S3 Appendix.}
\label{si_models}
{\bf Models parameterisation.} 

\paragraph*{S4 Appendix.}
\label{si_overcompensation}
{\bf Density-dependent mortality and overcompensation.} 

\section*{Acknowledgments}
This work has been funded under the project PesTwin, which has received funding from Cascade funding calls of NODES Program, supported by the MUR - M4C2 1.5 of PNRR funded by the European Union - NextGenerationEU (Grant agreement no. ECS00000036).

\nolinenumbers

% Either type in your references using
% \begin{thebibliography}{}
% Text
% \end{thebibliography}
%
% or
%
% Compile your BiBTeX database using our plos2015.bst
% style file and paste the contents of your .bbl file
% here. See http://journals.plos.org/plosone/s/latex for 
%\begin{thebibliography}{10}
%\end{thebibliography}

\bibliography{export}

%\subfile{sections/S1_inheritance}

%\subfile{sections/S2_agents}

%\subfile{sections/S3_models}

%\subfile{sections/S4_overcompensation}

\end{document}